\documentclass[a4paper,noshowpacs,noshowkeys,floatfix,twocolumn,preprintnumbers,nofootinbib]{revtex4-2}
\usepackage{color,amsmath,amsfonts,graphics,epsfig,amssymb}
\usepackage{multirow}
\usepackage{url}
\usepackage{hyperref}
\makeatletter
\renewcommand{\p@subsection}{}
\renewcommand{\p@subsubsection}{}
\makeatother
\begin{document}
\title{
{~ \hfill \rm INR-TH-2023-018}\\
~\\
The origin of high-energy astrophysical neutrinos: new results and prospects}
\thanks{To be published in \textit{Physics Uspekhi}, Russian
doi:10.3367/UFNr.2023.04.039581, English
doi:10.3367/UFNe.2023.04.039581.}
\author{Sergey Troitsky} \email{st@ms2.inr.ac.ru}
\affiliation{Institute for Nuclear Research of the Russian Academy of Sciences, 60th October Anniversary Prospect 7a, Moscow 117312, Russia}
\affiliation{Physics Department, M.V. Lomonosov Moscow State University, Leninskie Gory 1--2, Moscow 119991, Russia}
\begin{abstract}
High-energy neutrino astrophysics is rapidly developing, and in the last two years, new and exciting results have been obtained. Among them are the confirmation of the existence of the diffuse astrophysical neutrino flux by the new independent Baikal-GVD experiment, the discovery of the neutrino emission of our Galaxy, new confirmations of the origin of a part of astrophysical neutrinos in blazars, and much more. This brief review, based on the author's presentation at the session of the RAS Physical Science Division ``Gamma quanta and neutrinos from space: what we can see now and what we need to see more'', summarizes the results obtained since the publication of the review \cite{ST-UFN}, and can be
considered as a companion to it.
\end{abstract}
\maketitle

\thispagestyle{empty}
\newpage
\setcounter{page}{1}
\tableofcontents

\section{Introduction}
\label{sec:intro}
The study of astrophysical high-energy (TeV to PeV) neutrinos is presently at the stage of intensive development. The largest for today, IceCube neutrino telescope has amassed enough statistics to make conclusions about astrophysical sources of neutrinos, while Baikal-GVD and KM3NeT quickly increase their volumes and have started to produce their first data. However, in view of the new results, questions about the origin of these neutrinos are more abundant than answers.

Here, we attempt to summarize numerous new (published after Ref.~\cite{ST-UFN}, that is, in 2022--2023) results in the field of high-energy neutrino astrophysics, as well as long-term plans for the field's development. A broader review of the subject, and of the results obtained up to and including 2021, can be found in~\cite{ST-UFN}. A large part of the results mentioned in \cite{ST-UFN} are not discussed here, and references to them are not duplicated in order not to clutter up the present paper.

\section{Experimental news}
\label{sec:exp}
Let us first focus on significant advances in experiments that detect high-energy neutrinos. The results on astrophysical sources will be discussed in the following sections.
\subsection{Baikal--GVD}
\label{sec:exp:GVD}
Baikal-GVD, the largest neutrino telescope in the Northern Hemisphere, continues to increase its effective volume by gradually adding new clusters of optical modules (as of 2023, 12 clusters are operating, one of which is partially complete). Also, additional strings located in space
between clusters were added to the configuration of the experiment. They should work to improve registration efficiency and accuracy of determination of the neutrino parameters. In 2022, the first results of the experiment working in an incomplete configuration (2018--2021) were published.

A principal achievement, not only of the Baikal experiment, but also of the entire neutrino astronomy, was the confirmation of the very existence of high-energy astrophysical neutrinos. All previous studies of the neutrinos were based on results of one experiment, IceCube, which is not free, like any other, from systematic uncertainties. Based on the analysis of two samples of Baikal-GVD cascade neutrino events with the highest probability of astrophysical origin, the hypothesis of the absence of astrophysical neutrino flux was rejected \cite{Baikal-diffusePRD} with a statistical significance of $3.05\sigma$. The first sample included 16 events with reconstructed energies above 70~TeV (the highest energy was 1200~TeV). For the second one, the lower energy limit of 15~TeV was used, but only events with arrival directions from below the horizon were selected, which significantly reduced the atmospheric background. Eleven such events were recorded, of which two had energies above 70~TeV and therefore were included in both samples. The below-horizon event with the highest energy, 225~TeV, arrived from a remarkable direction in the sky, see below in Sec.~\ref{sec:extragal:blazars}.

Recall that the standard parametrization of the spectrum of the isotropic diffuse flux of one-flavor neutrinos and antineutrinos with a power-law function is given by 
\begin{equation}
\frac{dF_{\nu+\bar\nu}}{dE} = \Phi_{0}
\left(\frac{E_{\nu}}{\mbox{100~TeV}} \right)^{-\gamma}
\!\!\!
\times 10^{-18}~\mbox{GeV}^{-1}\,\mbox{cm}^{-2}\,\mbox{s}^{-1}\,
\mbox{sr}^{-1}.
\label{Eq:plaw}
\end{equation}
Traditionally, the analysis assumes equal fluxes of neutrinos of different flavors, so the total flux is obtained by multiplying Eq.~(\ref{tab:plaw-fits}) by three.  Parameters $\Phi_{0}$ and $\gamma$, obtained by Baikal-GVD, are given in
Table~\ref{tab:plaw-fits},
\begin{table*}
\begin{center}
\begin{tabular}{cccc}
\hline
\hline
Analysis & Energies & $\Phi_{0}$ & $\gamma$ \\
\hline
Baikal-GVD, upgoing cascades \cite{Baikal-diffusePRD}& 15--100~TeV &
$3.04^{+1.52}_{-1.27}$ & $2.58^{+0.27}_{-0.33}$ \\
IceCube combined \cite{IceCube-ICRC2023-combined}&
2.5~TeV--6.3~PeV& $1.80^{+0.13}_{-0.16}$ & $2.52\pm0.04$ \\
IceCube, starting tracks \cite{IceCube-ICRC2023-starting}&
3--550~TeV& $1.68^{+0.19}_{-0.09}$ & $2.58^{+0.10}_{-0.09}$ \\
\hline
\hline
\end{tabular}
\end{center}
\caption{\label{tab:plaw-fits}
Parameters of power-law fits (\ref{Eq:plaw}) of 
the diffuse astrophysical neutrino spectrum from the 2022-2023 analyses. }
\end{table*}
which can be considered as a supplement to Table~3 of Ref.~\cite{ST-UFN}. There, parameters obtained in two new IceCube analyses, discussed below, are also presented.

\begin{figure}
\centerline{\includegraphics[width=\columnwidth]{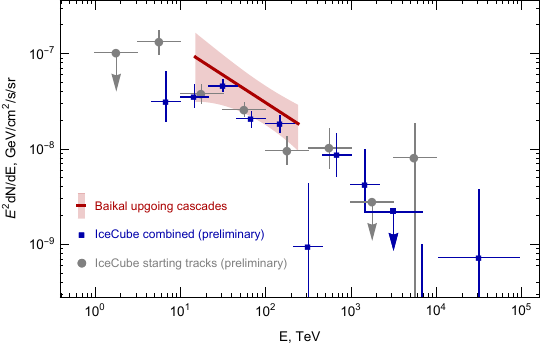}}
\caption{\label{fig:diff-fluxes}
Spectra of diffuse astrophysical neutrinos (one flavor, the sum of neutrinos and antineutrinos), obtained in 2022-2023 analyses.}
\end{figure}

Figure~\ref{fig:diff-fluxes}
demonstrates that the spectrum obtained in the Baikal-GVD analysis agrees well with IceCube's results. This result is important not only because it is obtained by a different team's independent analysis, but also because the two experiments differ significantly in sensitivity to different regions of the sky (Southern and Northern Hemispheres) and in methodology (ice and liquid water, see \cite{ST-UFN} for details).

\subsection{Experiments in the Mediterranean Sea: ANTARES and
KM3NeT}
\label{sec:exp:Mediterranean}
Another group of experiments detect neutrinos in liquid water
of the Mediterranean Sea. The ANTARES experiment has completed its multi-year operation in 2022, when the working volume of its coming replacement, the large KM3NeT detector ARCA, exceeded the volume of ANTARES. Final publications based on the full ANTARES dataset are expected in near future, some of the results are mentioned in Sec.~\ref{sec:extragal:blazars} and \ref{sec:gal:discovery}. Although KM3NeT already exceeds ANTARES by its volume, it is still small compared to IceCube and Baikal-GVD and has not yet reached sufficient exposure to detect the astrophysical diffuse neutrino flux. The experiment is increasing its working volume; in the end of 2022, it started to produce its first astrophysical results.

\subsection{IceCube}
\label{sec:exp:IceCube}
The IceCube experiment operates since 2008, and the main astrophysical results on high-energy neutrinos are still based on its data. In addition to increasing the data set, the IceCube team is working to improve the quality of event reconstruction and the accuracy of determination of neutrino energies and arrival directions. The new reconstructions are expected to account for ice properties more precisely. Machine learning techniques start to be applied for astrophysical analyses at the stage of reconstruction of individual IceCube events.

At the ICRC2023 conference, the IceCube experiment presented two new preliminary analyses of diffuse fluxes of astrophysical neutrinos, see Table~\ref{tab:plaw-fits} and Fig.~\ref{fig:diff-fluxes}. In particular, for the first time, a spectrum was constructed \cite{IceCube-ICRC2023-combined}, which combines the information from different observing channels, including cascades and tracks, selected and processed in different ways. In contrast to previous analyses, the quality of the fit of the spectrum by a broken power-law function is slightly better than that by a single power law.

Another new spectrum~\cite{IceCube-ICRC2023-starting} is based on the analysis of the tracks starting in the detector. This method of selection eliminates atmospheric muons efficiently, although, of course, it leaves atmospheric neutrinos in the sample. To isolate the latter contribution, statistical methods are used. This allows one to advance to lower energies in the astrophysical flux estimation. Note some discrepancy with the combined spectrum at low energies, see Fig.~\ref{fig:diff-fluxes}. When the starting-track spectrum is fitted with a broken power law, it even produces a break in the opposite direction, although this effect is not statistically significant. The problem of disagreement of the spectra obtained in different analyses, discussed in detail in \cite{ST-UFN}, retains its relevance.

Recently, several publicly available data sets related to the arrival directions of IceCube events have been released.
\begin{itemize}
 \item
\textit{Catalog of alert and ``alert-like'' track events,
IceCat--1}~\cite{IceCat}. Following the publication of its well-known result \cite{IceCube-TXS0506gamma} related to the coincidence of a high-energy neutrino event with the gamma-ray flare of the blazar TXS 0506+056, IceCube revised criteria for selecting and reconstructing public alerts, which are issued to inform the world's observatories about high-probability astrophysical neutrinos. The first-generation alerts were published in 2016--2018, the new ones -- starting from 2019. The catalog~\cite{IceCat} presents some of these new alerts together with results of the reprocessing of earlier events with the same new procedure. Events which satisfy the new alert criteria were selected from the old data, so that a homogeneous sample of 275 events was obtained. An important innovation is the inclusion of a veto related to the triggering of a surface-mounted unit that allows one to exclude certain events which have a high probability of being atmospheric. The catalog includes events from May 2011 through December 2019 (the experiment has been running since 2008). It is presumed that the information on newer events will be added to the online version of the catalog.
\item
\textit{Updated arrival directions of high-energy starting events (HESE)} \cite{IceCube-newHESEdirections}. 
This is another reprocessing of the entire dataset using a new reconstruction procedure that should take into account the properties of ice in the IceCube detector volume in a more correct way. Best-fit arrival directions and their (irregularly shaped) uncertainty regions in the sky are given for 164 events.
\item
\textit{Map of the Northern sky constructed from track events}. The numerical likelihood function used in the work on search for neutrinos from the NGC~1068 galaxy (\cite{IceCube-NGC1068}; for details, see Sec.~\ref{sec:extragal:Seyfert}), was presented. As in previous similar works, this function, defined on the celestial sphere, is related to the probability of detecting a local source of astrophysical neutrinos in this direction. It accounts both for the number of events coming from this direction and for their energies (the higher are the energies, the higher is the probability of the astrophysical neutrino origin).
\item
\textit{Sky map of cascade events}.
A similar likelihood function has also been published in connection with the observation of the neutrino emission from the Galactic plane (\cite{IceCube-gal-Science}, see section~\ref{sec:gal:discovery}). Cascade events were used for its construction.
\end{itemize}
Being openly accessible, these data can be utilized by researchers not affiliated with IceCube for new analyses and hypothesis testing (see, however, Sec.\ \ref{sec:old-problems} and \ref{sec:extragal:disapp}).

\subsection{New data, old challenges}
\label{sec:old-problems}
Application of new, refined methods of statistical analysis of raw data and of reconstruction of track and cascade events leads to a significant reduction of statistical uncertainties of the arrival directions, and for cascades -- also of the neutrino energy. With these developments, it becomes more and more clear that the accuracy of reconstruction of the neutrino properties is limited by systematic uncertainties. This manifests itself, in particular, in the differences in directions and energies obtained for the same events using different reconstructions (see e.g.\ illustrations in~\cite{ST-UFN}). For the IceCube experiment, one of the main sources of the uncertainty is the lack of knowledge of properties of the ice, that is of the medium in which the detected signal is formed and propagates.
Recently, IceCube publications started to present descriptions of how the systematic errors are taken into account in the data and to discuss directions of reducing these uncertainties \cite{Lagunas2021, Lagunas2023, IceCube-TDE-ICRC2023}. Presently, arrival directions of IceCube events with a high probability of the astrophysical origin are obtained using a simplified algorithm. Only for one event \cite{IceCube-event-for-which-syst}, the reconstruction was performed under assumptions of different models of ice properties. Obtained for this particular event, the systematic error was translated, following certain rules \cite{Lagunas2021}, to all neutrino alerts. The use of this procedure was motivated by the fact that multiple repetitions of simulations with different assumptions about the ice properties, even for a small amount of the most interesting events, required too much computer resources. Relatively recently, the full simulations were performed for several events, and has predictably shown that the actual reconstruction uncertainty due to insufficient knowledge of ice properties may be either smaller or larger than that estimated with the simplified method \cite{Lagunas2021}. The IceCube team is working on a solution to this problem~\cite{Lagunas2023}.

Development of an efficient approach for estimating systematic uncertainties in reconstruction of individual events remains a task for the future. Current published properties of IceCube events, including those in the catalog~\cite{IceCat}, were obtained in the simplified way described above. For practical purposes, additional systematic error can be accounted for by artificially increasing the statistical uncertainty~\cite{neutradio1, Resconi-blazars}. Thanks to the greater homogeneity of liquid water, compared to ice, and to the relative technical simplicity of controlling its properties, systematic uncertainties are expected to be less significant for other detectors. However, the same challenges remain relevant for all instruments.

Issues related to modeling uncertainty become very serious in the context of the increasing use of machine-learning techniques for event reconstruction; for more details see Sec.~\ref{sec:extragal:disapp}.

\section{Extragalactic neutrinos}
\label{sec:extragal}
Most probably, a large part of astrophysical high-energy neutrinos come from extragalactic sources.
New data and analyses confirm the origin of a significant fraction of high-energy astrophysical neutrinos in blazars (Sec.~\ref{sec:extragal:blazars}). At the same time, other extragalactic sources (Sec.~\ref{sec:extragal:Seyfert}, \ref{sec:extragal:TDE}), as well as our Galaxy (Sec.~\ref{sec:gal}), also contribute to the neutrino flux. 

\subsection{Neutrinos from blazars}
\label{sec:extragal:blazars}
Recall that blazars are powerful active galactic nuclei with relativistic jets directed at the observer. The radiation produced in the jet has a larger intensity  for the observer because of the relativistic effects, and this puts blazars among the brightest sources of non-thermal radiation in the Universe. The most universal marker of a relativistic jet directed to the observer is provided by the synchrotron radiation of relativistic electrons at the parsec scales, visible in the radio band with a very-long baseline radio interferometry (VLBI). Not all blazars are gamma-ray sources, although among extragalactic sources of high-energy gamma rays, they constitute the dominant population.

\subsubsection{Highest-energy neutrinos}
\label{sec:extragal:blazars:200+}
\begin{figure}
\centerline{\includegraphics[width=\columnwidth]{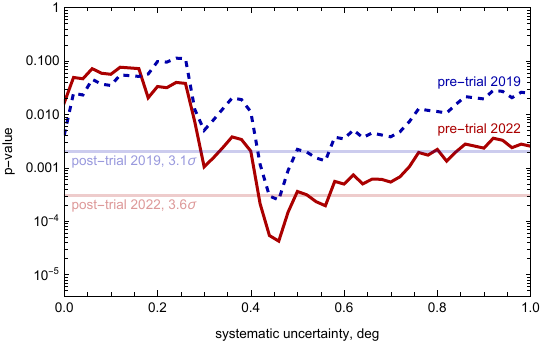}}
\caption{\label{fig:neutradio2022}
Probability of the null hypothesis of a random coincidence of high-energy neutrinos with radio blazars for data sets before 2019 and before 2022 for various estimated values of the additional systematic error (pre-trial). Horizontal lines indicate the significance obtained with the account of the choice of this value (post-trial).}
\end{figure}

\paragraph{New neutrino events: a direct test of the hypothesis.}
Statistical relationship between IceCube neutrino events and the population of blazars from the VLBI-selected sample was found in Ref.~\cite{neutradio1} for events with energies above 200 TeV, the data on which were published before and including 2019. The events collected in 2020--2022 have been analyzed by the same method in Ref.~\cite{neutradio2022}. The same event selection criteria and the same procedure of the analysis, established in Ref.~\cite{neutradio1}, were used. Fifteen new events have been added to the sample of 56 neutrinos used in \cite{neutradio1}. The statistical significance of the association of neutrinos with energies above 200~TeV with blazars increased from $3.1\sigma$ to $3.6\sigma$ (see Fig.~\ref{fig:neutradio2022}).
This result is a direct confirmation of the results of \cite{neutradio1}; the proportion of new blazar associations \cite{neutradio2022} among the neutrinos is consistent with that expected from \cite{neutradio1}.

\begin{figure}
\centerline{\includegraphics[width=0.83\columnwidth]{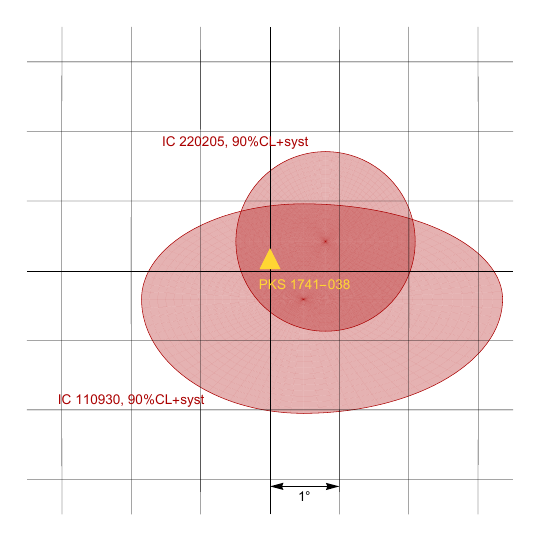}}
\centerline{\includegraphics[width=0.83\columnwidth]{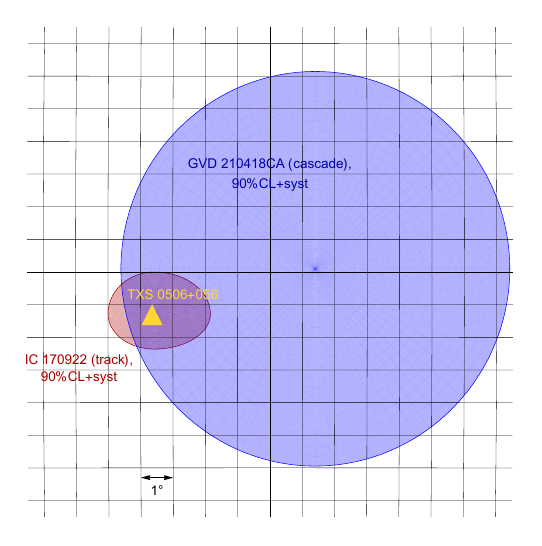}}
\centerline{\includegraphics[width=0.83\columnwidth]{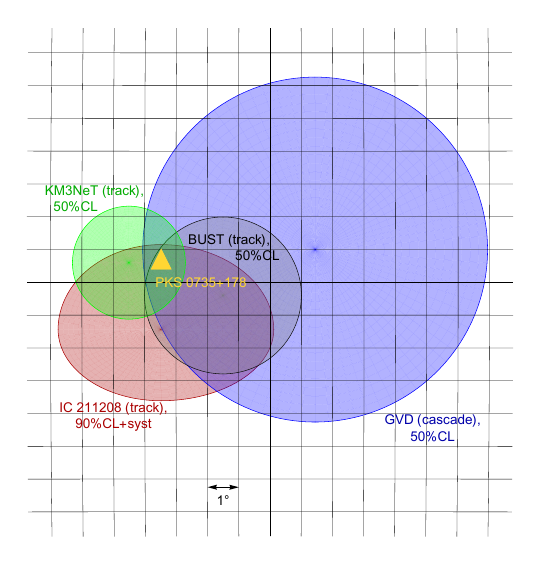}}
\caption{\label{fig:doublets}
Arrival-direction error contours for neutrinos associated with blazars (from top to bottom) PKS~1741$-$038, TXS~0506$+$056,
PKS~0735$+$178 (see text). }
\end{figure}

\paragraph{Repeated neutrinos from the same blazars.}
Accumulation of IceCube statistics, as well as the start of new experiments, Baikal-GVD and KM3NeT, resulted in observations of several neutrino events with the arrival directions coinciding with one and the same blazar. Let us focus on a few notable cases (see Fig.~\ref{fig:doublets}).

\textit{PKS~1741-038.}
One of the most powerful radio blazars in the sky, it was highlighted \cite{neutradio1} among the four most likely sources of neutrinos, based on the coincidence with the IC110930 neutrino event\footnote{Event identifiers indicate the experiment, IC -- IceCube, GVD -- Baikal-GVD, and the date of detection. For track events, only the most probable value of the neutrino energy is given, errors of determination of which are huge \cite{IceCube-TXS0506gamma,ST-UFN}.} and on high flux density of the radio emission from a compact component. In 2022, from the same direction, another neutrino, IC220205, has arrived that satisfied all the selection criteria used in \cite{neutradio1, neutradio2022}.

\textit{TXS~0506+056.}
The association of this blazar with the event IC170922 ($\sim 290$~TeV) started \cite{IceCube-TXS0506gamma} the history of observational associations between blazars and neutrinos. Both this neutrino event and this blazar were included in the sample of Ref.~\cite{neutradio1}. At the end of 2022, the Baikal-GVD experiment reported~\cite{Baikal-0506} a coincidence  of the cascade event GVD210418CA with the same blazar. This is the event with the highest energy ($225\pm 75$~TeV) among the cascades coming from directions below the horizon, registered by Baikal-GVD in 2018--2021. The estimated probability of its astrophysical origin is 99.67\%. Although the uncertainty of the arrival direction of this event ($6.0^{\circ}$, 90\% CL) is significantly larger than that of IceCube tracks, it is several times smaller than that of cascades in ice. This allows one to speak about the beginning of the neutrino point-source astronomy in the cascade channel.

\textit{PKS~0735+178.}
This source has attracted attention after the IC211208 event ($\sim$171~TeV). Although the event did not meet the selection criteria established in \cite{neutradio1} (the energy was below 200~TeV) and therefore has not been used in any statistical analysis, it coincided with an historical blazar flare, recorded in all bands, from radio to gamma rays. During the days of the December 2021 flare, neutrino events were recorded by all high-energy neutrino telescopes operating on Earth -- in addition to IceCube, these are Baikal-GVD~\cite{Baikal-ATel-0735, Baikal-paper-0735}, Baksan Underground Scintillation Telescope (BUST)~\cite{BUST-ATel-0735, BUST-paper-0735}, and KM3NeT~\cite{KM3Net-ATel-0735}. Neutrinos from this blazar have been discussed, in particular, in \cite{0735-int-1, 0735-int-4, 0735-VERITAS}. It is worth noting that, although BUST has small geometric volume and detects tracks of upgoing muons from neutrinos with an energy threshold of 1~GeV, i.e., orders of magnitude lower than those discussed here, its effective volume quickly grows with the neutrino energy, and the registration of one event per flare in BUST is consistent \cite{BUST-paper-0735} with one IceCube event, assuming a steeply falling neutrino spectrum of the source.

\paragraph{Other samples of IceCube events or blazars.}
As it was discussed in~\cite{ST-UFN}, statistically significant associations of high-energy neutrinos with blazars were also found in Ref.~\cite{Resconi:2020} that used selection criteria for events and blazars different from \cite{neutradio1}. The sample of neutrino events from \cite{Resconi:2020} (2009--2019) was again used in Ref.~\cite{Kun:2022} to find associations with sources, selected by fluxes in various bands, mostly blazars. The results of this study confirm the overall trend, including neutrino connection to flat-spectrum radio sources selected by the flux at 8~GHz. Unlike in other studies, the most efficient selection criterion of blazars on the basis of VLBI data was not used here: the sample was selected from the CRATES catalog, constructed on the basis of observations without the use of radio interferometry. About 4\% to 42\% of neutrinos in the sample \cite{Resconi:2020} can be associated with CRATES selected blazars (90\% CL interval), which is consistent with other estimates.

In 2023, the updated catalog of IceCube events having a high probability of astrophysical origin, IceCat-1, was published \cite{IceCat}. It contains events from 2011 through 2019, to which new reconstruction algorithms, currently used for alerts, were retroactively applied. Reconstructed arrival directions and energies of previously published events changed, in some cases significantly; many new events have also been added. Using this catalog for the analysis similar to \cite{neutradio1}, with direct application of the criteria from \cite{neutradio1}, results in the decrease \cite{IceCube:corr-IceCat} of the significance of associations between neutrinos and radio blazars compared to the original work, see discussion below in Sec.~\ref{sec:extragal:disapp}.

\subsubsection{All observed neutrinos}
\label{sec:extragal:blazars:TeV+}
\paragraph{Radio blazars as TeV to PeV neutrino sources.}
The relation between lower-energy neutrinos and radio blazars from the same VLBI-selected sample was established statistically \cite{neutradio2} on the basis of the published likelihood-function map containing generalized information about the arrival directions of all IceCube track events for the first 7 years of work. It was demonstrated that about 25\% of the flux of astrophysical muon neutrinos is associated with the population of blazars with a radio bright compact component. Separate events have been published for the 10-year dataset in another reconstruction. For this latter catalog, and using another analysis procedure, no directional correlations between neutrinos and blazars were found \cite{Zhou}, so that the upper limit on the fraction of neutrinos from blazars was set at $<30\%$. This is in agreement with \cite{neutradio2}; see also the discussion in the \cite{PlavinICRC2021}, and Sec.~\ref{sec:extragal:disapp} below.

Therefore, statistical analyses indicate that neutrinos, both of the highest and of somewhat lower energies, are related to one and the same population of blazars. The number of blazars in the sample is large, and individual sources are quite different. To understand the mechanisms of neutrino production, it is important to identify whether neutrinos with a broad energy spectrum are born in the same blazars, or different source populations are responsible for different neutrino energy ranges. Some progress in addressing this issue comes from Ref.~\cite{flares}, which established a statistically significant ($3.6\sigma$) correlation between matching a high-energy neutrino from the sample \cite{neutradio2022} to a radio blazar and the presence of additional lower-energy neutrinos arriving from the same direction at the same time ($\pm 1$~day). It is likely that the same blazars can produce neutrinos of significantly different energies.

\paragraph{Analyses based on other samples.}
The relation between \textit{all} events, registered by large neutrinos telescopes, with blazars is being confirmed using other data. Specifically, the ANTARES collaboration obtained \cite{ANTARES-blazars} indications (significance $2.2\sigma$) of the presence of a spatial correlation of arrival directions of events registered during the operation period of the experiment, with the same catalog of radio blazars that was used in \cite{neutradio1, neutradio2, neutradio2022}. In addition to this, they also conducted a search for neutrino flares from the directions of these blazars. The most notable was the excess of events from the blazar PKS~0242$+$1101 in 2013: a high-energy neutrino registered by IceCube came at the same time from the same direction, and the blazar was experiencing a powerful flare both in gamma rays (Fermi LAT) and in the radio band (OVRO).

Other selection criteria, primarily based on the optical spectra, formed the basis of the BZCAT catalog of blazars used in \cite{Buson1, Buson2}. Although this sample is not complete by any criterion (in particular, sources are distributed unevenly across the sky), it is expected to have a very low level of contamination by non-blazar objects. Ref.~\cite{Buson1} considered track events from the Southern sky, i.e. those which, for IceCube, come from directions above the horizon. The likelihood function map based on 7 years of IceCube data was used -- the same as, for the Northern sky, was used in~\cite{neutradio2}. The level of atmospheric muon contamination for events coming above the horizon is very high, so the maps for the Northern and Southern skies were constructed in different ways: for the Southern part, higher weight was given to the most energetic events, for which the probability of the astrophysical origin is higher. In~\cite{Buson1}, a rather involved method was applied in which ``hot spots'', that is several directions with high probability of the location of the neutrino source, were first identified in the Southern neutrino sky, and then those directions were tested for the presence of blazars. The association of neutrino ``hot spots'' with blazars was established at the confidence level of $4.7\sigma$. As for the Northern Sky, the same authors used \cite{Buson2} a different dataset, namely the updated likelihood map published along with the result \cite{IceCube-NGC1068}. The same conclusion was reached but with a significance of $2.7\sigma$. Note that using one and the same dataset for the Southern and Northern skies does not bring significant results for the latter \cite{Disappearing-corr}
(see Sec.~\ref{sec:extragal:disapp} for more details).

\paragraph{Theoretical implications.}
Production of neutrinos with energies ranging from TeV to PeV in the same sources, radio-bright blazars, is probably related to the interaction of protons, accelerated to energies about 20 times the energies of neutrinos, with X-ray photons (for more details see \cite{ST-UFN}). The maximal probability of the birth of neutrinos with $\sim$PeV energy corresponds to the interaction of protons with photons of ultraviolet light, but for lower energies of neutrinos, hard X-ray photons are required. Thus, observational results suggest that neutrinos are born in those parts of blazars, where photons with energies in a wide range are sufficiently abundant. On the other hand, the connection between the neutrino detection and the flux of a compact radio source, monitored by VLBI, indicates that neutrinos are produced within a few parsecs from the central black holes of blazars.

Not many mechanisms have been described in the literature that satisfy these requirements, the most stringent of which is the presence of a sufficient amount of X-ray photons. Hard X-ray radiation of blazars is often associated with Compton scattering of relativistic electrons on the photons of their own synchrotron radiation. The latter is exactly the kind of radio emission that is responsible for the VLBI flux, so it has been suggested \cite{neutradio2} that the region of the neutrino production coincides with the region where the compact radio emission comes from. Further development of this mechanism allowed one to construct~\cite{cores} a realistic two-zone model in which protons are accelerated near black holes, but interact with Compton photons in the so-called millimeter blazar core, an area near the base of the jet that gives the main contribution to the VLBI flux in the millimeter band \cite{Marscher}. In such a mechanism, the neutrino flux from a single blazar turns out to be relatively small, which agrees well with the estimates of the number of high-energy neutrino sources and with the lack of correlations between neutrinos and high-energy gamma-ray emission from blazars, discussed in \cite{ST-UFN}.

One of the predictions of models in this class is the large flux of hard X-ray emission of those blazars which are associated with neutrinos, in particular with those of the highest energies. This prediction has recently received an observational confirmation~\cite{neutxray}.

\subsection{Neutrinos from Seyfert galaxies}
\label{sec:extragal:Seyfert}
Another isolated individual neutrino source was associated with the M~77 galaxy, aka NGC~1068. This galaxy combines signs of nuclear activity (Seyfert type 2) and intense star formation. In \cite{IceCube-NGC1068}, the IceCube collaboration rejects the hypothesis of the absence of neutrino association with this galaxy at a statistical significance of $4.2\sigma$. Interpretation of this value is not straightforward because of the fact that, by itself, the significance of a neutrino hot spot in the direction close to NGC~1068 in the full-sky scan is~\cite{IceCube-NGC1068} $2.0\sigma$. The significance increases if the scan across the whole sky is replaced by a catalog of 110 ``probable sources'' compiled following rather arbitrary rules (for a more detailed discussion of this approach, see \cite{ST-UFN}). IceCube has been using such catalogs for some time, but it should be noted that the list has been significantly expanded when moving from the 2016 analysis \cite{IceCube7yrSources} to the 2019 analysis
\cite{IceCube-list2019}. In particular, in Ref.~\cite{IceCube-list2019}, NGC~1068 has been added to the list, along with 7 other galaxies with intense star formation, and in the very same paper, an excess of events from this direction in the full-sky scan was first detected. Since the events from 2011 to 2020 were used in \cite{IceCube-NGC1068}, the use of the term ``a priori fixed catalog'' in the context of the NGC~1068 source, included in 2019, is perhaps not fully justified.

Another unexpected difference between this and other sources, a very soft spectrum, is worth noting. It is well known that isolating the contribution of astrophysical neutrinos from the atmospheric background is possible only statistically, and is based on the distributions in the zenith angle and, most importantly, the energy (see \cite{ST-UFN}). The main part of atmospheric events has a soft spectrum with the power-law index $\approx 3.7$, while for astrophysical neutrinos, one expects the values of the index between about 2.0 and 2.7. In the analysis of events from NGC~1068, the value of the power-law index of $3.2\pm 0.2$ was obtained \cite{IceCube-NGC1068}, that is, the astrophysical origin of the neutrino exces is deduced mainly from the concentration of their arrival directions in a small region of the sky, not from their high energies.

Seyfert galaxies, as well as galaxies with intense star formation, are numerous, and the natural question arises about the contribution of other representatives of the same source classes to the neutrino flux. There are, for example, other similar nearby galaxies -- are they neutrino sources? This question is explored in Ref.~\cite{SemikozSeyfert}. There, a list of nearby galaxies similar to NGC~1068 was constructed and their expected neutrino luminosities were estimated. Taking into account these luminosity estimates and positions in the sky, it was found that the current IceCube's sensitivity should be sufficient to detect neutrino signal only from two more galaxies (besides NGC~1068 itself), NGC~4151 and NGC~3079. The authors analyzed a 10-year public catalog \cite{IceCube10yrDataPaper} of IceCube events and found excesses of neutrinos from these sources with statistical significances of $3.0\sigma$ and $3.9\sigma$, correspondingly. Note that, according to~\cite{SemikozSeyfert}, both these sources also have a soft spectrum, though the values of power-law indices are not given in the paper.

\subsection{Neutrinos from tidal disruption events}
\label{sec:extragal:TDE}
In the frameworks of one of observational follow-up programs triggered by IceCube neutrino alerts, a coincidence was found \cite{TDE1} of a high-energy IceCube event with an optical flare, probably associated with a tidal destruction of a star by the gravitational field of a supermassive black hole in the center of one of galaxies. Shortly afterward, another similar coincidence was found \cite{TDE2}. In both cases, comparison of optical and infrared observations of the suspected source revealed a delay in the infrared flash that can be explained by scattering of radiation on large amounts of dust. This motivated a statistical study \cite{TDEcorr2021} in which a catalog of similar events was constructed and the third match was found. Statistical significance of the coincidence of 3 of the 40 high-energy IceCube events, included in the sample, with the flares associated with accretion of matter on supermassive black holes, is $3.6\sigma$. In interpreting this quantitative result, it should be taken into account that two events, which motivated the consideration of this sample, were included in the calculation of the significance.

\subsection {``Disappearing'' correlations}
\label{sec:extragal:disapp}
As it was noted above, changing the IceCube event reconstruction procedure often leads to a noticeable change in the key characteristics of the neutrinos, energies and arrival directions. It can be noticed that some of the interesting associations of neutrinos with potential astrophysical sources become less statistically significant when newly reprocessed samples of the same events are used. Let us focus on a few of these cases.

\paragraph{TXS 0506$+$056.}
Although active galactic nuclei have been proposed as potential sources of high-energy neutrinos long before the discovery of the latters, the most serious attention to the possible connection between neutrinos and blazars was caused by the publication of two IceCube papers \cite{IceCube-TXS0506gamma,IceCube-TXS0506-flare} dedicated to the same source, a fairly ordinary blazar TXS 0506$+$056. Ref.~\cite{IceCube-TXS0506gamma} announced the observation of a gamma-ray flare of this blazar a few days after detecting a high-energy neutrino from the direction of this
source in September 2017. Ref.~\cite{IceCube-TXS0506-flare}, published simultaneously, discussed a neutrino flare in 2014 from the same direction, found in a subsequent analysis of old data. The statistical significance of the 2014 flare, according to \cite{IceCube-TXS0506-flare}, was $4.0\sigma$ before the account of the penalty factors associated with the selection of analysis options, and $3.5\sigma$ after accounting for them. In the new reconstruction used by IceCube in 2023, changes in the characteristics of the same events led to a decrease in the first value down to $3.3\sigma$ \cite{IceCube-TXS0506-ICRC2023} (only events prior to October 2017 were used). This means that, taking into account the trial corrections, the flare described in~\cite{IceCube-TXS0506-flare} has the significance of $2.7\sigma$ in the new reconstruction. The best-fit neutrino flux of this possible flare has decreased twice \cite{IceCube-TXS0506-ICRC2023} compared to the original publication \cite{IceCube-TXS0506-flare}.

As for the coincidence of a high-energy event to the same blazar's gamma-ray flare, its statistical significance would also be diminished in the present-day analysis, though for a different reason: a large number of newer alert events did not result in the detection of any flare of a coincident source. Such an analysis, however, is not very easy to perform correctly due to the change in alert criteria after \cite{IceCube-TXS0506gamma}, mentioned in Sec.~\ref{sec:exp:IceCube}.

\paragraph{Blazar populations.}
A similar story develops with the effects found in the analyses of blazar populations. As we noted in Sec.~\ref{sec:extragal:blazars:200+}, the significance of associations between IceCube neutrinos with energies above 200~TeV and VLBI blazars decreases \cite{IceCube:corr-IceCat} when the new reconstruction \cite{IceCat} is used, compared to Ref.~\cite{neutradio1}, which used the originally published arrival directions and energies. Here one can also recall Ref.~\cite{Zhou}, where no significant correlation of the same blazars with lower-energy neutrinos was found in one of newer IceCube reconstructions, see the discussions in~\cite{PlavinICRC2021, ST-UFN}.

A recent paper~\cite{Disappearing-corr} is devoted entirely to comparing correlations of all IceCube neutrinos with  blazar populations in the two datasets, the 7-year~\cite{IceCube7yrSources} and 10-year~\cite{IceCube10yrDataPaper} ones. The authors of \cite{Disappearing-corr} use the method of hot spots in the sky map, applied in~\cite{Buson1} to the likelihood map of the 7-year data set in the Southern Sky. For the 10-year set of events, they built a similar map themselves \cite{Disappearing-corr}. The statistical significance of the result~\cite{Buson1} has deteriorated from $4.7\sigma$ to $0.3\sigma$ with the transition to the reconstruction of Ref.~\cite{IceCube10yrDataPaper}. The catalog of VLBI blazars used in \cite{neutradio2} was also examined in the same way. It's interesting that it was found \cite{Disappearing-corr} to correlate with ``hot spots'' of the Southern Sky at a significance level of $3.2\sigma$ (the original Ref.~\cite{neutradio2} used the Northern sky map and another method) -- but only in the 7-year dataset. Use of the updated reconstruction \cite{IceCube10yrDataPaper} results in a dilution of this newly found effect as well.

\paragraph{Tidal disruption events.}
In Sec.~\ref{sec:extragal:TDE}, we discussed three IceCube events which coincided with episodes of intense accretion on supermassive black holes \cite{TDEcorr2021}. IceCube's recent work \cite{IceCube-TDE-ICRC2023} utilizes a new catalog of neutrino events to test the association with tidal disruption events (TDE) selected by criteria similar to \cite{TDEcorr2021}. Preliminary results indicate that there is no statistically significant correlation. It is noted that for two out of three neutrinos previously associated with TDE, the arrival directions changed in the new reconstruction in such a way that TDEs are now outside of the error contours, while the third event is no longer included in the sample at all.

\paragraph{Possible causes.}
For an external observer, it is difficult to judge why associations with astrophysical sources and their populations,  significant in earlier analyses, consistently disappear when newer reconstructions for the same IceCube events are used. One potential explanation that is reached, for example, in \cite{Disappearing-corr}, is that all of these sources do contribute to the neutrino flux, but their contributions are smaller than it seemed from the initial tests. This explanation is certainly possible, but it would look more natural in the case when the effect weakens with new data, not with reprocessing of the same data by new algorithms. Indeed, in statistical analyses at the limit of the sensitivity, weak effects tend to open up as positive fluctuations, and therefore are often not visible in the next set of data\footnote{Just in the same way as positive fluctuations lead to unusual successes in sports and other human activities, which the person cannot repeat afterwards, see, e.g., \protect\cite{Kahneman}.}. But it is not easy to imagine -- though, with a small probability, it is possible -- that in a number of different analyses, statistically significant results appeared due to the use of incorrect reconstruction, and the improvement of the latter led to their blurring.

One possible reason for the decrease of significance with the transition to newer reconstructions is in a delicate balance of statistical and systematic uncertainties. Event reconstructions, developed in recent years, widely use machine learning techniques. This leads to a firmly proven significant reduction in statistical errors of determination of parameters of the particle that triggered the detector. At the same time, systematic uncertainties often give rise to serious, and difficult to control, problems. The point is that machine learning involves a training dataset with known characteristics, on which, in fact, the algorithm learns to determine the characteristics of other events. In modern high-energy astroparticle physics, characterization of the primary particle is possible only indirectly, since for energetic photons, neutrinos and charged cosmic particles, only products of their interaction with the detector or the atmosphere is registered. Therefore, as training datasets, one has to use artificial sets of events based on Monte-Carlo simulations. The latter involve detailed modeling of the processes occurring in the detector, and thus requires their perfect knowledge. In the case of IceCube, ideal knowledge of optical properties of ice in the volume of a cubic kilometer is not yet possible, and therefore the training datasets are necessarily built using certain assumptions. If these assumptions are not fully correct, a new method may reduce statistical errors in determining the neutrino arrival direction, but the central value may be shifted due to the training dataset's non-ideality. In this case, simpler, but less model-dependent analyses, would win, like those used at the initial stages of the experiment's operation. It is difficult to understand, even for those directly involved in the data processing, to what extent can this situation take place in reality. It is clear that the final criterion for testing hypotheses about the origin of astrophysical neutrinos should be related to analysis of data of independent experiments, primarily those using liquid water, for which both statistical and systematic errors are considerably smaller \cite{ST-UFN} (see Sec.~\ref{sec:prospects:water} below). So far, this work is just beginning \cite{Baikal-0506, ANTARES-blazars,Baikal-Nikita}.

\section{Galactic neutrinos}
\label{sec:gal}
Interactions of cosmic rays with matter in the Galactic disk result in the neutrino production, and this process guarantees a certain flux of high-energy neutrinos from the disk of the Milky Way. In addition, individual Galactic objects, in which cosmic rays are accelerated and interact with ambient matter, are potential sources of Galactic neutrinos. Some contribution is also expected from interactions of cosmic particles leaving the Galaxy with circumgalactic gas. Possible sources of Galactic neutrinos and the history of their search are described in the reviews \cite{ST-UFN, Kheirandish:gal}. In 2022, this search succeeded. 

\subsection{Discovery of the Milky-Way neutrino emission}
\label{sec:gal:discovery}
Ref.~\cite{neutgalaxy} analyzed the distribution of arrival directions of IceCube events which had a high probability of astrophysical origin, energies above 200~TeV and good quality of reconstruction, in the absolute value of the Galactic latitude, $|b|$. The median value of $|b|$ for the set of 71 events, selected according to the criteria established in previous analyses, is $|b|_{\rm med}\approx 21^{\circ}$. The expected value of $|b|_{\rm med}$ for simulated sets of events distributed in the sky according to the IceCube exposure is $\langle |b|_{\rm med}\rangle \approx 36^{\circ}$, and the probability to obtain $|b|_{\rm med}\le 21^{\circ}$ as a result of a random fluctuation is $4\times 10^{-5}$ (statistical significance of $4.1\sigma$). This established the concentration of neutrino arrival directions from this sample to the Galactic plane. Similar analysis for all events from the catalog~\cite{IceCube10yrDataPaper} demonstrated that the Galactic component is also observed at lower energies~\cite{neutgalaxy}. Making use of a single observable $|b|_{\rm med}$ is the simplest, uncertainty-free way to search for the Galactic component of the neutrino flux, since it does not require any assumptions about specific sources and no parameter tuning.

The ANTARES collaboration also studied~\cite{ANTARES-ridge} the distribution of arrival directions of events and took advantage of the fact that for this experiment, central regions of the Galaxy are observed below the horizon, which makes it possible to significantly reduce the atmospheric background from this direction. The traditional ``on--off'' method was used, in which a narrow rectangle $4^{\circ}\times 60^{\circ}$ in the center of the Galaxy was chosen as the signal (``on'') area. The background was determined by the number of events in ``off''-areas with the same observational conditions, but in the directions away from the Galactic center. An indication to the excess of events from the signal area with the statistical significance of $2.0\sigma$ was found. Both track and cascade events were used in the study, but the main contribution to the signal comes from tracks: the number of neutrino tracks with energies above 1~TeV in the ``on'' region was 21 with the background expectation of $11.7 \pm 0.6$, and the number of cascades was 13 with the background of $11.2 \pm 0.9$.

Finally, in 2023, a paper was published by the IceCube collaboration \cite{IceCube-gal-Science}, which presented results of the search for the signal from the Galactic plane in the set of cascade events. The choice of the cascade channel was motivated by the high background for the tracks from the Galactic-center direction, which, for IceCube, is always observed above the horizon. The search methodology used in this study was substantially different. In fact, it was not about finding an arbitrary signal from the Galaxy, but about testing three specific models of diffuse radiation associated with interactions of cosmic rays with gas in the disk. The first of the models was based on Fermi-LAT observations of diffuse gamma rays in the GeV range: it was assumed that this radiation was associated with decays of $\pi^{0}$ mesons, and neutrinos come from decays of $\pi^{\pm}$ mesons born in the same interactions. The distribution of arrival directions and the spectrum (power law with the 2.7 exponent) were extrapolated from the GeV range of the Fermi LAT to the TeV range of IceCube. The other two models, called KRA$\gamma_{5}$ and KRA$\gamma_{50}$, were obtained by modeling the propagation of cosmic particles in the Galaxy  and their interactions with matter with the DRAGON code; the difference between the two is the energy of the assumed cosmic-ray spectral cutoff (5 or 50~TeV, respectively). For each of the three models, templates of the expected distribution of neutrinos in directions and energies were derived, taking into account the detection and selection procedures for IceCube events. Then these expected distributions were compared, with the help of the likelihood function, with those actually observed. The null hypothesis of the absence of the neutrino flux from the Galactic plane was excluded with statistical significance of $4.71\sigma$ ($\pi^{0}$ template), $4.37\sigma$ (KRA$\gamma_{5}$ template), and $3.96\sigma$ (KRA$\gamma_{50}$ template). The final significance, taking into account these three trials, is $4.5\sigma$.

The three independent results described above allow us to speak with confidence about the discovery of high-energy Galactic neutrinos: the Milky Way is now visible in the neutrino sky. However, as we will see below, reliable conclusions about the origin of these neutrinos are still far away. It would be of  interest to verify the obtained results with the Baikal-GVD data. The first, so far few, published data on cascade events do not contradict the assumption of a Galactic component. Moreover, the arrival-direction error circles of 3 out of 11 Baikal-GVD events with energies above 100~TeV overlap, and this triplet is close to the Galactic plane \cite{Baikal-Nikita}. This area of the sky is quite interesting; it contains, among other possible sources, one of a few Galactic binary systems observed in the gamma-ray band, LSI~$+61~303$, as well as the point of the maximum of the likelihood function, used by IceCube to search for point sources in the Northern sky based on the 7-year track data \cite{IceCube7yrSources}. In more recent analyses, the maximum of the likelihood shifted to the direction close to NGC~1068, see Sec.~\ref{sec:extragal:Seyfert}.

\subsection{Comparison of analyses}
\label{sec:gal:compare}
\begin{figure}
\centerline{\includegraphics[width=\columnwidth]{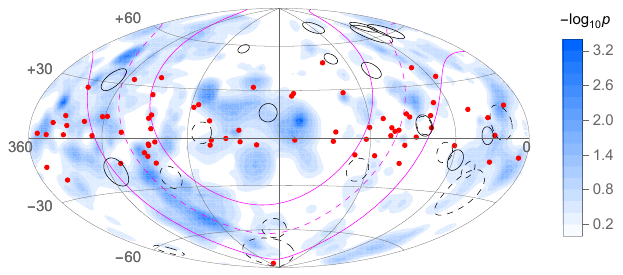}}
\caption{\label{fig:sky-gal}
Sky map (equatorial coordinates, Hammer projection) with the distribution of the likelihood function constructed for IceCube cascade events \cite{IceCube-gal-Science}. Arrival directions of 71 IceCube tracks used in the analysis of Kovalev et al.~\cite{neutgalaxy} are shown as red dots. For the arrival directions of Baikal-GVD cascade events from Ref.~\cite{Baikal-Nikita}, 90\% CL uncertainty contours are given (solid lines -- the sample of $E>100$~TeV events, dashed lines -- upgoing cascades; one event present in both samples and coinciding with TXS~0506$+$056, see Sec.~\ref{sec:extragal:blazars}, is shown by a double contour). The dashed purple line is the Galactic plane, and the solid purple lines bound the band $|b|<20^{\circ}$.}
\end{figure}

Fig.~\ref{fig:sky-gal} presents the sky map which shows the likelihood function used in the IceCube cascade analysis
\cite{IceCube-gal-Science}, arrival directions of IceCube track events from the sample used in \cite{neutgalaxy}, and Baikal-GVD cascade events \cite{Baikal-Nikita}.
One can observe some concentration of neutrinos to the broad band near the Galactic plane in all cases. However, a direct comparison of the results of the different tests is hardly possible. Indeed, analyses~\cite{neutgalaxy, ANTARES-ridge, IceCube-gal-Science} refer to different regions of the sky and different neutrino energies, and most importantly, they use fundamentally different approaches, see Table~\ref{tab:gal-3anal}.
\begin{table*}
\begin{center}
\begin{tabular}{cccc}
\hline
\hline
Analysis & Energies & Method & Significance\\
\hline
Kovalev et al.~\cite{neutgalaxy}&$\gtrsim 200$~TeV & median $|b|$, tracks
 & $4.1\sigma$ \\
ANTARES~\cite{ANTARES-ridge}&$\sim 1-100$~TeV & on/off, tracks and cascades &
$2.0\sigma$\\
IceCube~\cite{IceCube-gal-Science}&$\sim 1-100$~TeV & distribution templates, cascades
& $4.5\sigma$ \\
\hline
\hline
\end{tabular}
\end{center}
\caption{\label{tab:gal-3anal}
Analyses (2022-23) in which high-energy neutrinos from our Galaxy were detected.}
\end{table*}

\begin{figure}
\centerline{\includegraphics[width=\columnwidth]{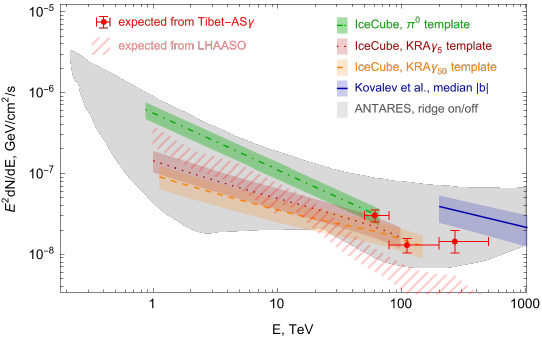}}
\caption{\label{fig:gal-nu-spec}
Estimated total, sky-integrated Galactic neutrino flux as a function of energy, according to the results of Kovalev et al.~\cite{neutgalaxy}, ANTARES~\cite{ANTARES-ridge, ANTARES-ICRC2023-G1}, and IceCube~\cite{IceCube-gal-Science}. The hatched area and the points with error bars represent neutrino spectra expected from measurements of Galactic diffuse gamma rays by LHAASO~\cite{LHAASODiffuseGal} and
Tibet-AS$\gamma$~\cite{TibetDiffuseGal}, respectively (see 
Sec.~\ref{sec:gal:gamma}). }
\end{figure}

To compare the Galactic neutrino spectra obtained under different
assumptions, one can use their generalized characteristic, the full-sky flux of neutrinos of the Galactic origin, assessed on the basis of assumptions about its part contributing to the detected flux. This is most easy to do when the flux is searched using a template, like it was done in the IceCube paper \cite{IceCube-gal-Science}. This recalculation for the ANTARES~\cite{ANTARES-ridge} result was presented in \cite{ANTARES-ICRC2023-G1}. For the work~\cite{neutgalaxy}, it can be easily derived from the fraction of the diffuse neutrino flux attributable to the Milky Way, estimated in the paper. The results of comparison of the spectra recalculated in this way are shown in Fig.~\ref{fig:gal-nu-spec}.
They indicate qualitative agreement of all three analyses, but do not allow for detailed quantitative comparisons without reference to specific data used to obtain them. We will return to this discussion in Sec.~\ref{sec:gal:width}.

\subsection{Galactic diffuse neutrino and gamma-ray emission above 100~TeV}
\label{sec:gal:gamma}
Simultaneously with the diffuse neutrino emission, photons of the same energy range should be produced in the Galactic disk. Unlike extragalactic, this gamma radiation, with energies about tens of TeV, is not strongly absorbed due to the $e^{+}e^{-}$ pair production on background photons. Therefore, the photons, accompanying the neutrino emission from the Milky Way, can be detected. They were indeed discovered by the Tibet-AS$\gamma$ experiment \cite{TibetDiffuseGal} even earlier than the Galactic neutrino emission was. In 2023, Galactic diffuse gamma rays were detected also by an independent experiment, LHAASO~\cite{LHAASODiffuseGal}. Overall, the results of the two experiments are qualitatively consistent with each other, though the flux measured by LHAASO is formally somewhat lower than that obtained by Tibet-AS$\gamma$. This may be related to particular details of accounting for the contribution of Galactic point sources by the two experiments, or to other systematic uncertainties.

Although the fluxes, $F_{\nu}$ and $F_{\gamma}$, of neutrinos and photons, born simultaneously in high-energy proton-proton interactions, are roughly related by a simple law,
$$
F_\nu(E_\nu)\approx 2F_\gamma(E_\nu/2)
$$
(see discussion in~\cite{ST-UFN}), a practical comparison of results of neutrino experiments with predictions based on this formula are not easy. The point again is that the analyses refer to different areas of the sky and to different energies, and are obtained by different methods. Some attempts of such a comparison were made in~\cite{neutgalaxy, nu-gamma-gal1,nu-gamma-gal2, nu-gamma-gal3}; they indicate a good overall agreement between neutrino~\cite{neutgalaxy, ANTARES-ridge,IceCube-gal-Science} and photon~\cite{TibetDiffuseGal,LHAASODiffuseGal} Milky-Way diffuse fluxes, thereby indirectly confirming the origin of both in hadronic interactions. This is illustrated in Fig.~\ref{fig:gal-nu-spec}, where we present, together with neutrino fluxes, their expectations from the diffuse photon emission from the Galaxy measured by the two experiments. For the latter, we used recalculation from Ref.~\cite{nu-gamma-gal2}.

\subsection{Galactic neutrino angular distribution}
\label{sec:gal:width}
To quantitatively understand the origin of Galactic neutrinos, the key issue is the spatial distribution of the neutrino sources. Here, at first glance, there seems to be some discrepancy between the results of different analyses. However, as we will see in a moment, this discrepancy is more apparent than real.

The angular distribution of diffuse Galactic emission in templates, used by IceCube~\cite{IceCube-gal-Science}, is to a large extent determined by the distribution of matter in the disk of the Galaxy and therefore follows a narrow band in the sky (a few degrees wide, like the visible Milky Way). At the same time, a model-independent study~\cite{neutgalaxy} indicates the excess of events in a much wider band $|b| \lesssim 20^{\circ}$. Note that the main result~\cite{neutgalaxy} does not use, nor directly predicts, the width of the band, and this value appears only in the supplementary analysis, both for the main sample of events with energies above 200~TeV and at somewhat lower energies.

To understand the reasons for these different results, let us note a significant scatter in the normalizations of the  Galactic diffuse neutrino spectra obtained by IceCube~\cite{IceCube-gal-Science} using different templates, see Fig.~\ref{fig:gal-nu-spec}. At low energies, where the main statistics are accumulated, the difference in fluxes, determined assuming different templates, reaches several standard deviations. The highest statistical significance is obtained for the template which uses extrapolation by three orders of magnitude without involving any quantitative physical model. The discrepancy in the results, obtained under different assumptions, may indicate that at least some of these assumptions are wrong. At the same time, due to poor angular resolution for IceCube cascade events, it is not possible to determine the shape of the Galactic neutrino signal in a reliable manner without the use of an a priori fixed template. Figure~\ref{fig:sky-gal} qualitatively demonstrates that the excess of IceCube cascade events from the Galactic plane is possibly consistent with a wider distribution of arrival directions, than the templates assume.

Any model of the origin of the diffuse neutrino flux from interaction of cosmic particles with matter in the Galactic disk makes use both of the distribution of this matter and of the spatially-dependent spectra of the cosmic radiation. While the gas distribution is fairly well known from observations, cosmic-ray concentration and spectra in remote regions of the disk can only be obtained indirectly. The reason for this is the complex motion of charged particles in magnetic fields, which in addition are poorly known themselves. To date, many models of the propagation of charged cosmic particles in the Galaxy are based on simplistic assumptions. One of the key assumptions is that the spectrum of Galactic cosmic rays, recorded in the vicinity of the Earth, is representative for the Galaxy. There are a number of indications that such an assumption does not hold, see e.g.\ \cite{Semikoz-bubble1, Semikoz-slope1, GiacintiSemikoz2023,Semikoz-ANTARESgal}. In particular, the presence of a nearby source of cosmic rays, combined with higher gas density in the so-called Local Bubble, can lead to an increased contribution of the nearby Galaxy region to the observed neutrino flux~\cite{Semikoz-bubble1, Semikoz-bubble2}. Projected on the celestial sphere, this flux would come from higher Galactic latitudes than the main contribution of the disk, which may lead to broadening of the latitude distribution of Galactic neutrinos.

\section{Prospects}
\label{sec:prospects}
Enormous background of non-astrophysical events, both atmospheric neutrinos and muons, together with large statistical and systematic uncertainties in the determination of neutrino parameters, remain the main factors limiting further development of neutrino astrophysics towards identifying and studying the sources of high-energy neutrinos. It is not surprising that the future of this field hinges on overcoming these two challenges.

\subsection{Combating atmospheric backgrounds: high energies and high statistics}
\label{sec:prospects:large}
Since the atmospheric background is unavoidable, and each individual atmospheric neutrino is not different from an astrophysical one, advances in separating the astrophysical signal can only be linked to the increase in the number of detected events, which requires a large effective volume of the detector. On the one hand, higher statistics allows for more precise separation of the contribution of astrophysical neutrinos to the total observed flux, because ensembles of neutrinos of atmospheric and astrophysical origin have different distributions in energies, flavor composition, and arrival directions \cite{ST-UFN}. On the other hand, increasing the volume of the detector makes it possible to detect rare events with very high energies, for which the atmospheric background is low.

Among specific plans for the construction of new experiments, substantially larger than those currently in operation, is the IceCube-GEN2 project~\cite{IceCube-Gen2}. It is proposed to expand the existing IceCube detector (1~km$^{3}$) up to the instrumented volume of 7.9~km$^{3}$ (mainly due to the increase in area, because at large depth, the optical properties of ice deteriorate considerably). The amount of neutrino events, compared to IceCube, should increase, approximately, proportionally to the volume. Outside of the present detector, the distance between the strings of optical modules will be significantly increased.

In addition to IceCube-GEN2, projects of detectors with very large volume include TRIDENT~\cite{TRIDENT-8km3} and HUNT~\cite{HUNT-30km3} installations which are discussed below in Sec.~\ref{sec:prospects:water}.

For neutrinos with energies $\gtrsim 10^{17}$~eV, the atmospheric background is absent, and low fluxes become the main problem. Main hopes in this energy range are related to the detection of neutrinos by the radio emission of the cascade processes they cause (ARA~\cite{ARA}, ARIANNA~\cite{ARIANNA}, RNO-G~\cite{RNO-G}, GRAND~\cite{GRAND} projects, etc.). Possible fluxes of astrophysical neutrinos of even higher energies are so small  that to record them, one needs a spacecraft observing large volumes of the Earth's atmosphere (JEM-EUSO~\cite{JEM-EUSO}, POEMMA~\cite{POEMMA}, etc.). Discussion of these energy ranges is beyond the scope of this paper.

\subsection{Fight for accuracy: detectors in liquid water}
\label{sec:prospects:water}
The IceCube Upgrade project~\cite{IceCube-upgrade} (not to be confused with IceCube-GEN2) will soon be implemented at the South Pole. Among other things, it will include the installation of additional calibrating devices that will allow for more precise control of optical properties of the ice and thus for improving the accuracy of neutrino event reconstruction. However, ice properties vary within the operating volume, while calibration will only be carried out in its small part. Main prospects of refining the determination of neutrino properties are associated with the use of detectors in liquid water, see e.g.\ a discussion in Ref.~\cite{ST-UFN}.

To date, the largest liquid-water neutrino detector is Baikal-GVD~\cite{Baikal-GVD}, whose volume as of 2023 is about 0.6~km$^{3}$, and is increasing by about 0.1~km$^{3}$ every year. The data obtained from Baikal-GVD in incomplete configuration are being used for astrophysical analyses since 2018, see previous sections. Other $\sim 1$~km$^{3}$ scale detectors include KM3NeT~\cite{KM3Net}
(Mediterranean Sea), which began data collection in 2022, and projected instruments: P-ONE~\cite{P-ONE} (which will make use of oceanological infrastructure off the Pacific coast of Canada; work is underway on the prototype) and NEON~\cite{NEON-1km3-dense} (South China Sea, with a denser arrangement of optical modules compared to the current telescopes). Also in the South China Sea, it is proposed to place the TRIDENT~\cite{TRIDENT-8km3} experiment with the working volume of 8~km$^{3}$. Finally, the most far-reaching plans have been recently presented by a group of researchers associated with the above-mentioned LHAASO experiment: the HUNT~\cite{HUNT-30km3} project is aimed at construction of a neutrino telescope with the working volume of up to 30~km$^{3}$. For the location of such a huge instrument, one considers either a place far enough offshore in the South China Sea, or Lake Baikal, which is shallower, but has convenient infrastructure. Equipment tests at the first of these two sites were performed in 2022-23, and at Baikal they are scheduled for 2024. The neutrino telescope at Lake Baikal became, more than 25 years ago, the first to detect \cite{Baikal-1st-neutrinos1997,Baikal-1st-neutrinos1999} a neutrino event with the method subsequently used to obtain all the results discussed in this paper (see also the historical review~\cite{Spiering-UFN}). Maybe in another 25 years, the facility will become the world's most ambitious neutrino detector with the working volume 30 times larger than that of the present-day IceCube.

We noted above that a significant limitation of astrophysical neutrino experiments is related to systematic uncertainties, which are different for each experiment. Therefore, the key reliability factor of neutrino astrophysics is the unification of the efforts of different experiments, which use different methods and have different sensitivities to the Northern and Southern skies. Since 2013, these efforts are being developed~\cite{Spiering:2014,Spiering:GNN} within the framework of the Global Neutrino Network\footnote{\tt https://www.globalneutrinonetwork.org/.} (GNN). Probably, the creation of focused thematic working groups, which would include representatives of different experiments, will allow one to advance much further in understanding astrophysical neutrino sources. Positive experience of work of such groups is already seen in the field of ultra-high energy cosmic rays.

\section{Conclusions}
\label{sec:concl}
\begin{itemize}
 \item
High-energy neutrino astrophysics is entering a new stage of development with the start of cubic-kilometer scale experiments in liquid water, Baikal-GVD and KM3NeT. For the first time, the very existence of astrophysical neutrinos was confirmed independently of IceCube by the Baikal-GVD experiment.
 \item
Results of various analyses, including those formally verifying previously formulated hypotheses with new data, confirm the origin of a significant part of astrophysical neutrinos in blazars. Due to systematic differences between approaches and datasets, the fraction of the neutrino flux associated with blazars is hard to determine precisely.
 \item
There are strong indications that some of the neutrinos detected at the Earth are born in other extragalactic sources, among which are  Seyfert galaxies and centers of galaxies in which tidal disruption events occur.
 \item
The neutrino emission of the Milky Way has been detected. Three independent analyses, based on different data, are qualitatively consistent with each other and with the observations of diffuse Galactic gamma rays. The discrepancies in the results of model-dependent quantitative analyses point to the possible need to revise models of the cosmic-ray propagation in the Galactic disk.
 \item
The statistical significance of a number of statements indicating to point sources of neutrinos, including those related to the very first detected TXS~0506$+$056 source, is greatly reduced when using the same events re-processed with new IceCube reconstruction algorithms. This demonstrates the importance of understanding and correct accounting for systematic uncertainties in the experiment.
 \item
Projects of future neutrino telescopes are motivated by increasing the exposure, required to separate the astrophysical signal from the atmospheric background more precisely, and by the use of liquid water to reduce statistical and systematic uncertainties. The experiments of today's generation are combined in the Global Neutrino Network, where data sharing and collaborative analyses should help to eliminate a number of uncertainties in the conclusions already in the coming years.
\end{itemize}

The author is indebted for interesting and useful discussions on various topics, related to the origin of high-energy astrophysical neutrinos, to his colleagues, co-authors, and participants of the Scientific Session of PSD RAS ``Gamma quanta and neutrinos from space: what we can see now and what we need to see more''.

\bibliography{try23}
\bibliographystyle{apsrev4-1}

\end{document}